\relax
\documentclass[letterpaper]{article} 
\usepackage{aaai22}  
\usepackage{times}  
\usepackage{helvet}  
\usepackage{courier}  
\usepackage[hyphens]{url}  
\usepackage{graphicx} 
\usepackage{natbib}  
\usepackage{caption} 
\DeclareCaptionStyle{ruled}{labelfont=normalfont,labelsep=colon,strut=off} 
\usepackage{algorithm}
\usepackage{algorithmic}
\usepackage{amsmath}
\usepackage{mathtools}
\usepackage{changes}
\usepackage{hyperref}
\usepackage{multirow}
\usepackage{booktabs}
\usepackage{scrextend}
\deffootnote[1em]{1em}{1em}{\thefootnotemark.\space}
\usepackage{newfloat}
\usepackage{listings}
\floatstyle{ruled}
\newfloat{listing}{tb}{lst}{}
\floatname{listing}{Listing}
\title{Unbiased Pairwise Learning to Rank in Recommender Systems}

 \author{
 Yi Ren  \equalcontrib \thanks{henrybjren@tencent.com},
 Hongyan Tang \equalcontrib \thanks{violatang@tencent.com},
 Siwen Zhu \thanks{siwenzhu@tencent.com}
 }
 
 \affiliations{PCG, Tencent Inc.}

\usepackage{bibentry}

\begin{document}

\maketitle

\begin{abstract}
Nowadays, recommender systems already impact almost every facet of people’s lives. To provide personalized high-quality recommendation results, conventional systems usually train pointwise rankers to predict the absolute value of objectives and leverage a distinct shallow tower to estimate and alleviate the impact of position bias. However, with such a training paradigm, the optimization target differs a lot from the ranking metrics valuing the relative order of top-ranked items rather than the prediction precision of each item. Moreover, as the existing system tends to recommend more relevant items at higher positions, it is difficult for the shallow tower based methods to precisely attribute the user feedback to the impact of position or relevance. Therefore, there exists an exciting opportunity for us to get enhanced performance if we manage to solve the aforementioned issues. 

Unbiased learning-to-rank(LTR) algorithms, which are verified to model the relative relevance accurately based on noisy feedback, are appealing candidates and have already been applied in many applications with single categorical labels, such as user click signals. Nevertheless, the existing unbiased LTR methods cannot properly handle multiple feedback incorporating both categorical and continuous labels. Accordingly, we design a novel unbiased  LTR algorithm \footnote{The source code of the proposed algorithm is released at https://github.com/phyllist/ULTRA.} to tackle the challenges, which innovatively models position bias in the pairwise fashion and introduces the pairwise trust bias to separate the position bias, trust bias, and user relevance explicitly.  Experiment results on public benchmark datasets and internal live traffic show the superior results of the proposed method for both categorical and continuous labels.
    
\end{abstract}

\section{Introduction}
Recommender systems model user behaviors (e.g., click, purchase, share, and dwell-time) to learn user preference and recommend top-ranked items to the corresponding user. In industrial recommendation applications, pointwise learning is the mainstream preference modeling approach, which aims to learn the absolute probability of user behavior rather than the relative preference between items. Therefore, there is a gap between the training objective of pointwise learning and the actual ranking objective in recommendation scenarios. On the contrary, pairwise learning focuses on minimizing the number of inversions to predict relative orders between items, which often performs better in practice and has been widely applied to search engines \cite{pasumarthi2019tf} as predicting relative preference is more consistent with the nature of ranking.

However, user behaviors are often biased in recommendation scenarios. For instance, a user might click and watch a video simply because it is ranked high rather than it is the one that the user likes the most, so higher-ranked items often exhibit significantly larger CTR than lower-ranked items. Due to the position bias, it is difficult for pairwise learning to learn the accurate relative preference, which results in the rare application of pairwise learning in recommender systems. Thus, to estimate relative user preference accurately and improve ranking performance in recommender systems, it is essential to correct position bias from user behaviors.

To deal with position bias in recommender systems, prior works propose to add a shallow tower to model bias directly and alleviate its impact \cite{zhao2019recommending,guo2019pal}. However, bias features and relevance-related features are coupled together with the shallow tower based models, which makes it rather difficult to eliminate bias effectively \cite{wang2018position}. Unbiased learning to rank (LTR) methods \cite{ai2018unbiased,hu2019unbiased,agarwal2019addressing} model relation between the absolute values of unbiased relevance and the biased user feedback to learn the unbiased ranker, which perform well in search engines. Nonetheless, the existing methods can only work for categorical labels instead of continuous labels, as the relationship between relevance and user feedback in continuous tasks does not adhere to their core assumptions shown in Equation 3. Moreover, as continuous feedback (e.g., dwell time and video watch ratio) is essential to model user preference, it is critical to design an unbiased LTR method for continuous labels to learn unbiased relative preference accurately in recommender systems.

To achieve the goal, we propose a novel unbiased pairwise LTR method, which models position-based examination bias in the pairwise fashion and introduces the pairwise trust bias to better model the underlying bias and learn the relative preference more accurately. Compared with the state-of-the-art pointwise unbiased LTR methods, the proposed method models the correlation between the unbiased relevance pairs and the biased user feedback pairs directly, which bears no assumption mentioned above on labels and can work for both categorical and continuous labels. Moreover, pairwise debiasing is more consistent with the objective of ranking, which achieves better performance of debiasing and preference learning. Furthermore, the introduction of pairwise trust bias separates the position-dependent examination bias and the position-pair-dependent relative trust bias explicitly for better correction. Experiment results on benchmark datasets and online A/B testing show that the proposed method outperforms not only the SOTA ranking models in recommender systems but also the SOTA unbiased LTR methods commonly used in search engines.

The main contributions of the paper are as follows:

\begin{itemize}
\item We propose a novel unbiased pairwise LTR method for better bias correction and unbiased relative preference learning. To the best of our knowledge, the proposed method is the first unbiased LTR method that can work for both categorical labels and continuous labels.
\item We extend the regression-based EM algorithm \cite{wang2018position} to the pairwise setting to estimate the parameters of the proposed method effectively.
\item We conduct offline and online experiments to evaluate the performance of our method. Results on benchmark datasets show that our method outperforms both the SOTA unbiased LTR methods and the SOTA rankers in recommendation scenarios. Online A/B test in one of the world's largest content recommendation platforms shows a significant 2.08\% improvement of the business metric.
\end{itemize}

\section{Related Work}
In this section, we discuss related works on unbiased learning to rank and debiasing methods in recommender systems.
\subsection{Unbiased Learning to Rank}
Recently, unbiased LTR has been actively studied as a promising approach to learning from biased feedback. \cite{joachims2017unbiased} first present the counterfactual framework to learn the theoretically unbiased ranker via the inverse propensity weighting (IPW) estimator. To estimate the propensity, \cite{wang2016learning} and \cite{joachims2017unbiased} propose the methods of result randomization and intervention. To reduce the negative impact on user experience, some works propose to learn the propensity from biased feedback directly, such as DLA \cite{ai2018unbiased}, regression EM \cite{wang2018position} and Unbiased LambdaMART \cite{hu2019unbiased}. The position-dependent click noise \cite{joachims2017accurately} is ignored in these methods, which is addressed by \cite{agarwal2019addressing}. However, all of the above methods can only work for categorical labels because of their core assumption of equation 3. In contrast, our proposed method can work for both categorical and continuous labels.
\subsection{Modeling Biases in Recommender Systems}
There have been some works on modeling and eliminating biases in recommender systems.  \cite{zhao2019recommending} and \cite{guo2019pal} both propose to employ a shallow tower to estimate and alleviate the impact of bias by combining the output of the shallow tower and the main model with summation or multiplication. However, due to the coupling of bias features and relevance-related features, shallow tower based methods cannot work effectively. \cite{wu2021unbiased} apply the idea of \cite{wang2018position} to one recommendation scenario and learn unbiased ranker with IPW framework and regression-EM method. However, it still can only work for categorical labels, which is inadequate for typical recommender systems. The proposed method in this paper extends the IPW framework to the pairwise setting to deal with continuous labels in recommender systems, which achieves superior performance of debiasing and preference learning.

\section{Methodology }
In this section, we provide the general framework of unbiased learning to rank in recommender systems for both categorical and continuous labels. 
\subsection{Unbiased LTR for Categorical Labels}
Suppose there are $N$ positions, for a user request $u$, item $x_i$ is displayed at position $i  \!\! \in \!\! [1, N]$, $r_i$ is the unbiased relevance of  user-item pair $(u,x_i)$. For simplicity, we only consider binary relevance, and one can easily extend it to the multi-level case. Then we can calculate the risk function as follows:
\begin{equation}
R_{rel}(f) = \int_{}{}L(f(u,x_i), r_i) dP(r_i=1,u,x_i)
\end{equation}
where $f$ denotes a ranking system, $L(f(u,x_i),r_i)$ denotes the loss function based on performance metrics and $P(r_i \! \! = \!\! 1,u,x_i)$ denotes the probability distribution of positive $r_i$ on $x_i$. For notation simplicity, the position information of items is omitted from the loss function. As most performance metrics such as Discounted Cumulative Gain (DCG) \cite{jarvelin2002cumulated,jarvelin2017ir} and Average Relevance Position (ARP) \cite{wang2018lambdaloss} only consider relevant items, the risk function here is calculated only based on items with positive relevance label.

\begin{equation}
\hat{f}_{rel} = \arg\min_{f}  \sum_{u,x_i,r_i>0}{} L(f(u,x_i), r_i) 
\end{equation}

If the unbiased relevance label is available, we can learn a ranker by minimizing the empirical risk function shown in Equation 2. However, it is infeasible to collect enough true relevance labels in large-scale recommendation applications as the recommendation results are personalized and we cannot afford to find enough suitable judges for all users. Thus, abundant biased user actions (e.g., click, purchase, like) are valuable labels if we can fill the gap between user actions and relevance labels. As assumed in Position-Based Model (PBM) \cite{richardson2007predicting}, the user clicks the item if and only if he examines a relevant item, and the probability of examination only depends on the position and is independent with relevance. Let $e_i$ denotes whether item $x_i$ is examined by the user, $c_i$ denotes the user action label of $x_i$, we can model the relationship between $c_i$ and $r_i$ as follows:
\begin{align}
P(c_i=1|u,x_i,i)&=P(r_i=1|u,x_i,i) P (e_i=1|u,x_i,i) \nonumber  \\
&= P(r_i=1|u,x_i) P (e_i=1|i)
\end{align}
\label{eq3}

Then we can derive the risk function and empirical risk function based on user action label:
\begin{align}
R_{unbiased}(f) &= \int_{}{}  \frac{L(f(u,x_i), c_i)} {P (e_i=1|i)} dP(c_i=1,u,x_i,i)  \vspace{1ex} \nonumber \\
& = \int_{}{}  L(f(u,x_i), c_i) d \frac {P(c_i=1,u,x_i,i) } {P (e_i=1|i)} \nonumber \\
& = \int_{}{}  L(f(u,x_i), c_i)  dP(r_i=1,u,x_i) \nonumber \\
& = \int_{}{}  L(f(u,x_i), r_i)  dP(r_i=1,u,x_i) \nonumber \\
& = R_{rel}(f)
\end{align}
\begin{equation}
\hat{f}_{rel} = \hat{f}_{unbiased} = \arg\min_{f}  \sum_{u,x_i,i,c_i>0}{} \frac {L(f(u,x_i), c_i)} {P (e_i=1|i)}
\end{equation}

As shown in Equation 4, one can see that $R_{unbiased}$ with the inverse propensity weighting (IPW) \cite{joachims2017unbiased} loss equals to $R_{rel}$ in fact. In other words, if we can estimate the examination probability $P(e_i=1|i)$ accurately, we can learn an unbiased ranker based on user action labels.

Due to the assumption between the probability distribution of relevance labels and user action labels shown in Equation 3, the aforementioned method can only be applied to tasks with categorical labels rather than continuous labels, which tends to exhibit much more complex relationships between the user action labels and the true relevance. However, continuous labels such as dwell time and video watch ratio are important feedback of user satisfaction and are commonly used in industrial recommender systems. To correct position bias for continuous labels, we propose a novel unbiased LTR method which circumvents the previous assumption between the unbiased relevance labels and the biased user action labels. Details of the method would be provided in the following sections.

\begin{table}
\setlength{\belowcaptionskip}{-0.5cm}
\begin{tabular}{cl}
\hline
Notation & Description \\
\hline
$u$ & User request with user profile and context. \\
$x_i$ & Features of item displayed in position $i$. \\
$r_i$ & Unbiased recommendation relevance of $x_i$. \\
$c_i$ & Biased synthesized label of $x_i$. \\
$e_i$ & Examination of $x_i$. \\
$\theta_i$ & $P (e_i=1|i)$ \\
$\theta_i^-$ & $P (e_i=1|i,c_i=0)$ \\
$\epsilon_{ij}^+$ & $P(c_i>c_j | e_i=1,e_j=1,r_i>r_j,i,j)$ \\
$\epsilon_{ij}^-$ & $P(c_i>c_j | e_i=1,e_j=1,r_i<=r_j,i,j)$ \\
$\gamma_{u,x_i,x_j}$ & $P(r_i>r_j|u,x_i,x_j)$ \\
$\beta_{u,x_i}$ & $P(r_i>0|u,x_i)$ \\
\hline
\end{tabular}
\caption{Notations and Descriptions}
\end{table}


\subsection{Supervision Label Synthesis}
To provide high-quality recommendations, industrial recommender systems often consider multiple live metrics simultaneously \cite{tang2020progressive}. For example, we would like to maximize the user’s video view count, dwell time, and other satisfaction-related metrics (e.g., liking and sharing) in the video recommendation scenario. To directly model multiple live metrics simultaneously in one single model, we define a synthesized label based on diverse user actions. As shown in Equation 6, the synthesized label merges multiple user actions (e.g., click, dwell time, and satisfaction) through a combination function $\Phi$, where $M$ is the number of user actions considered in live metrics. The combination function is usually set based on business goals, and its form can be fairly flexible to capture complex relations between live metrics. For instance, weighted summation and weighted multiplication are commonly used in industrial recommender systems. 

\begin{equation}
c = \Phi(action_1, action_2, ... , action_M)
\end{equation}

There are two reasons to combine multiple user behaviors to the synthesized label for the learning objective. First, the gap between the training objective and the live metrics can be bridged through learning from the synthesized label, as the synthesized label is closely related to the ultimate business goals. Second, different from pointwise learning, pairwise learning and listwise learning care more about the relative order of the items rather than the absolute scores, which means that the absolute predicted scores of different unbiased LTR models might not be in the same scale. Thus, if we model different user actions through different unbiased LTR models, we could not combine predictions of different user actions in a theoretically sound way. 

\subsection{Unbiased LTR for Continuous Label}
Due to the continuous user feedback, the synthesized label is not categorical, and the aforementioned unbiased LTR method cannot work. To circumvent the limitation of the strong assumption between unbiased relevance labels and biased user action labels, we propose a novel unbiased LTR method from the perspective of pairwise learning, which can work for both categorical and continuous labels.

In the pairwise setting, the loss function is defined on item pairs. Let $\hat{y}_i=f(u,x_i)$ denotes the prediction for pair $(u,x_i)$, $L(\hat{y}_i,r_i, \hat{y}_j ,r_j)$ denotes the pairwise loss function and $P(r_i \!\!> \!\! r_j,u,x_i,x_j)$ denotes the probability distribution of positive relevance pair on item pair $(x_i,x_j)$, the risk function and the empirical risk function are defined in Equation 7 and Equation 8 respectively. Similar to unbiased LTR for categorical label, we also consider positive relevance pairs only as negative and neutral pairs should not be involved to optimize common ranking metrics such as DCG and ARP.

\vspace{-0.5cm}
\begin{equation}
 R_{rel}(f)  =    \int_{}{} L(\hat{y}_i,r_i, \hat{y}_j ,r_j) dP( r_i  >  r_j,  u,  x_i,  x_j ) 
\end{equation}
\vspace{-0.8cm}

\begin{equation}
\hat{f}_{rel}  =   \arg\min_{f}   \sum_{u, x_i, x_j,  r_i >  r_j}  L(\hat{y}_i,r_i, \hat{y}_j ,r_j)
\end{equation}

\subsubsection{Pairwise PBM based Unbiased LTR.}To model the relation between unbiased relevance pair  $(r_i,r_j)$ and biased synthesized label pair $(c_i,c_j)$, we extend the PBM to the pairwise setting. Specifically, we assume that the examination of item $x_i$ and $x_j$ are independent and if  $x_i$ and $x_j$ are both examined, $x_i$ exhibits larger synthesized label than $x_j$ if and only if $x_i$ is more relevant than $x_j$, which is:
\begin{equation}
P(c_i \! > \! c_j,e_i=1,e_j=1 | u,x_i,x_j,i,j) \! = \! \theta_i \theta_j \gamma_{u,x_i,x_j}
\end{equation}

Then we can derive the risk function based on the positive synthesized label pairs: 
\begin{align}
&\!\!\! \!  R_{unbiased}(f) \nonumber  \\
&\!\!\! \! \!\!= \!\! \int_{}{}  \frac {L(\hat{y}_i, \! c_i, \hat{y}_j, \! c_j)} {\theta_i \theta_j } \! P(  e_j \! \! = \! \! 1| \! u,\! i, \! j, \! x_i, \! x_j \!, \! c_i \!\!> \!\!c_j \!) dP( c_i \!\! > \!\! c_j, \! u, \! x_i, \! x_j, \! i , \!j ) \! \nonumber  \\
& \!\!\! \! \!\!= \!\! \int_{}{}  \frac {L(\hat{y}_i, \! c_i, \! \hat{y}_i, \! c_j)} {\theta_i \theta_j }  dP( c_i \!\! > \!\! c_j, \! u, \! x_i, \! x_j,i,j ,e_i=1,e_j=1\!) \! \nonumber  \\
&\!\!\! \! \!\!= \!\! \int_{}{}  \frac {L(\hat{y}_i, \! c_i, \! \hat{y}_i, \! c_j)} {\theta_i \theta_j }  dP( r_i \!\! > \!\! r_j, \! u, \! x_i, \! x_j,i,j ,e_i=1,e_j=1\!) \!  \nonumber  \\
&\!\!\! \! \!\!= \!\! \int_{}{}  \frac {L(\hat{y}_i, \! r_i, \! \hat{y}_j, \! r_j)} {\theta_i \theta_j }  dP( r_i \!\! > \!\! r_j, \! u, \! x_i, \! x_j,i,j) P(e_i \! = \! 1|i)P(e_j \!= \! 1|j) \! \nonumber  \\
&\!\!\! \! \!\!= \!\! \int_{}{}  {L(\hat{y}_i, \! r_i, \! \hat{y}_i, \! r_j)} dP( r_i \!\! > \!\! r_j, \! u, \! x_i, \! x_j \!) \! = R_{rel}(f)
\end{align}

Based on the justification that $R_{unbiased}$ is an unbiased estimator of $R_{rel}$, we can learn an unbiased ranker through minimizing the empirical risk function ${loss_{IPW}}$ shown in Equation 11, where $L_{ij}=L(\hat{y}_i, \! c_i, \! \hat{y}_j , \! c_j)$. It is worth noting that samples with positive labels and zero labels are separately handled explicitly for better precision in the empirical risk function, as the posterior probability of examination is different for samples with different labels. In detail, the posterior examination probability is definitely 1 for items with a positive label but is uncertain for items with zero labels.
\begin{align}
& \!\!\!\!\!loss_{IPW} = \!\!\!\!\!\!\!\!\!\!\!\!\!\!\!\!\!\!   \sum_{u, x_i, x_j,  i,   j,  c_i >  c_j,c_j>0}  \! \frac {L_{ij}} {\theta_i \theta_j} \! +  \!\!\!\!\!\!   \sum_{u, x_i, x_j,  i,   j, c_i > c_j,c_j=0}  \!\!\!\!  \frac {L_{ij}h_{ij}} {\theta_i \theta_j} \!,  \!\!
\end{align}
where $h_{ij}=P(e_j=1|u,x_i,x_j,i,j,c_i>c_j,c_j=0)$.



\begin{figure*}[!htbp]
\setlength{\belowcaptionskip}{-0.3cm}
	\centering
	\includegraphics[width=0.9\textwidth, trim=40 490 40 70,clip]{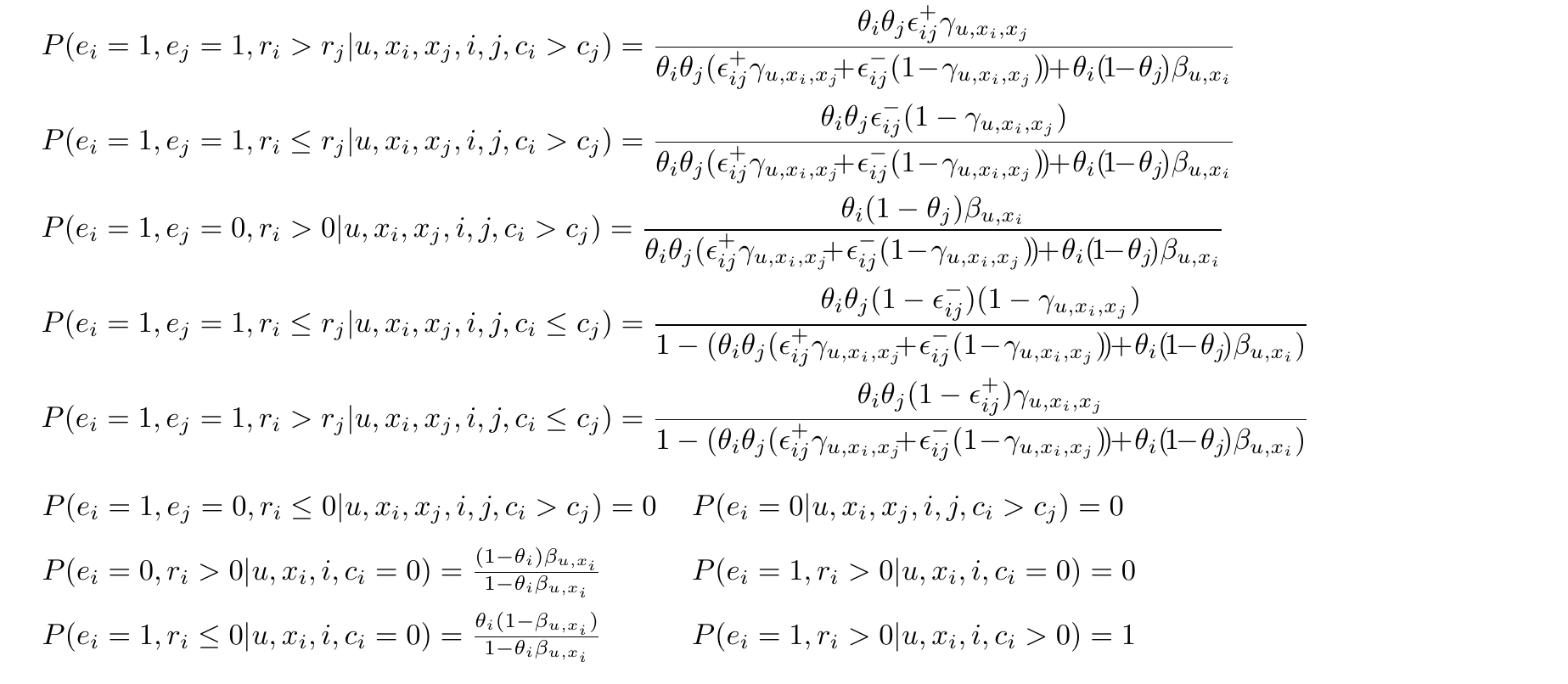}
	\caption{Formulas in Expectation Step}
	\label{estep}
\end{figure*}


\subsubsection{Pairwise Trust Bias based Unbiased LTR.} The pairwise PBM based unbiased LTR method assumes that if $x_i$ is more relevant than $x_j$ and they are both examined, $x_i$ will exhibit a larger synthesized label than $x_j$. However, according to the user study in \cite{joachims2017accurately}, users tend to trust higher-ranked items more, so higher-ranked items exhibit larger click ratios even if items in different positions are all examined and equally relevant, which is called the trust bias. To estimate the relation between unbiased relevance pair $(r_i,r_j)$ and biased synthesized label pair $(c_i,c_j)$ more accurately, we incorporate the trust bias explained above and extend it to the pairwise setting in this paper. Since the trust bias only depends on the exposed position, we formulate the pairwise trust bias in Equation 12 and Equation 13 and assume that $P(c_i \! > \! c_j|e_i=1,e_j=1, r_i>r_j,u,x_i,x_j,i,j)$ and $P(c_i \! > \! c_j|e_i=1,e_j=1, r_i \leq r_j, u,x_i,x_j,i,j)$ are independent of $(u,x_i,x_j)$. Specifically, for a positive relevance pair $r_i>r_j$, the probability of exposing a positive synthesized label pair $c_i >c_j$ is not 1 but $\epsilon_{ij}^+$, while a non-positive relevance pair also has the probability $\epsilon_{ij}^-$ to exhibit a positive synthesized label pair. $\epsilon_{ij}^+$ and $\epsilon_{ij}^-$ are only position-dependent and follows the constraint $0<\epsilon_{ij}^-<\epsilon_{ij}^+<1$.
\vspace{-0.1cm}
\begin{align}
P(c_i \! > \! c_j|e_i=1,e_j=1, r_i>r_j,i,j) \! = \epsilon_{ij}^+ \\ 
P(c_i \! > \! c_j|e_i=1,e_j=1, r_i \leq r_j,i,j) \! = \epsilon_{ij}^-
\end{align}


Then we can calculate the probability of a positive synthesized label pair. As shown in Equation 14, there are two cases to expose a positive synthesized label pair $c_i \!\! > \!\! c_j$. First, if $x_j$ is examined, the relation between $c_i$ and $c_j$ depends on the relation between $r_i$ and $r_j$ and the trust bias. Second, if $x_j$ is not examined, the positive pair $c_i \! > \! c_j$ reduces to the positive data point $c_i \! > \! 0$, which only depends on the examination and relevance of $x_i$ according to PBM.
\begin{align}
\!\!\!\! & P(c_i  >  c_j|u,x_i,x_j,i,j)   \nonumber  \\
&  =  P(c_i > c_j, e_i=1, e_j=1|u,x_i,x_j,i,j)  \nonumber  \\
& + P(c_i>0, e_i=1, e_j=0|u,x_i,x_j,i,j)  \nonumber \\
 & = \!  \theta_i \theta_j  ( \epsilon_{ij}^+ \gamma_{u,x_i,x_j} \!\!\! + \! \epsilon_{ij}^-(1 \! - \! \gamma_{u,x_i,x_j})\!) \!\! + \! \theta_i (1 \! - \! \theta_j\!) \beta_{u,x_i}
\end{align}

As we can no longer deduce the state of $r_i \! > \! r_j$ from the condition of $c_i \! > \! c_j,e_i \! = \! 1,e_j \!= \!1$, $R_{rel}$ could not be derived from $R_{unbiased}$ with the introduction of pairwise trust bias. To tackle this issue, we calculate the probability of positive relevance pair given the positive synthesized label pair and true examination, which is:
\begin{align}
\!\!\!\!\! & P(r_i \! > \! r_j|u,x_i,x_j,i,j,c_i>c_j,e_i=1,e_j=1)   \nonumber  \\
&  =\! \frac {P(c_i\!\! >\!\! c_j| e_i \! = \! 1,\! e_j \! = \! 1,\!r_i \!\!>\! \! r_j, \! u, \! x_i,\!x_j,\! i, \!j)P(r_i \!\! > \!\! r_j|u,\! x_i,\! x_j)} {P(c_i\! >\! c_j,\! e_i=1,\! e_j=1|u,x_i,x_j,i,j)}   \nonumber  \\
& = \! \frac {\epsilon_{ij}^+ \gamma_{u,x_i,x_j}} {\epsilon_{ij}^+ \gamma_{u,x_i,x_j}  +  \epsilon_{ij}^-(1  -  \gamma_{u,x_i,x_j})} = m_{ij}
\end{align}

Finally, we define a Bayes-IPW loss based on $loss_{IPW}$ and $m_{ij}$, which compensates and corrects for both the position-based examination bias and trust bias. One can see from Equation 16 and Equation 15 that the Bayes-IPW loss reduces to the $loss_{IPW}$ in the noisy free case of pairwise PBM, i.e., when $\epsilon_{ij}^+=1$ and $\epsilon_{ij}^--=0$.
\begin{align}
\!\!\!\!\!\!\!  loss_{Bayes-IPW} \! = \!  \arg\min_{f} \!\!\!\!\!\!\!   \sum_{u, x_i, x_j,  i,   j, \atop  c_i >  c_j,c_j>0} \!\!\!\!\!\!\!  \frac {L_{ij}m_{ij}} {\theta_i \theta_j} \! 
 +  \!\!\!\!\!\!\!\!\!  \sum_{u, x_i, x_j,  i,   j, \atop c_i > c_j,c_j=0}  \!\!\!\!\!\!\!\!\!  \frac {L_{ij}m_{ij}h_{ij}} {\theta_i \theta_j} \!,  \!\!
\end{align}
where ${h_{ij}}$ is calculated as follows:
\begin{align}
\!\!\!\!\!\!\!\!\!\! & h_{ij} = (e_j=1|u,x_i,x_j,i,j,c_i>c_j,c_j=0)  \nonumber  \\
& = \!\! \frac {P(e_j \!= \! 1|u, \! x_j,\!j,\! c_j \!\! = \! 0) P(c_i \!\! > \!\! c_j|u,\!x_i,\!x_j,\!i,\!j,\!e_j \! = \! 1,\! c_j \! = \! 0)} {P(c_i \! > \! c_j|u,x_i,x_j,i,j,c_j \! = \! 0)}  \nonumber  \\
& = \!\! \frac {\theta_i \theta_j^-  ( \epsilon_{ij}^ + \gamma_{u,x_i,x_j} \!\!\! + \! \epsilon_{ij}^-(1 \! - \!\! \gamma_{u,x_i,x_j})\!)}
 {\!  \theta_i \theta_j^- \!\!\ ( \epsilon_{ij}^+ \gamma_{u,x_i,x_j} \!\!\! + \! \epsilon_{ij}^-(1 \!\! - \! \gamma_{u,x_i,x_j})\!) \!\! + \!\! \theta_i (1 \!\! - \! \theta_j^- \! ) \beta_{u,x_i}}\!,
\end{align}
where $\theta_j^-$ denotes the posterior examination probability when the synthesized label of item $x_j$ is known to be 0.

\subsubsection{Direct Optimization for Ranking Metrics.}As pairwise loss \cite{burges2005learning} focuses on minimizing the number of pairwise errors and neglecting the relative importance of item pairs with different relevance, which does not match well with common performance metrics such as DCG and ARP. To directly optimize the ranking metrics in the model, we refine the loss function following the practice of LambdaRank \cite{burges2006learning}. The main idea of LambdaRank is to incorporate a delta NDCG to directly optimize the evaluation metric NDCG, where delta NDCG denotes the difference between NDCG scores if item  $x_i$ and $x_j$ are swapped in the ranking list \cite{burges2006learning,burges2010ranknet}.
As the unbiased true relevance is not available, we assume that the biased synthesized label is positively correlated to the unbiased relevance, and refine the pairwise loss $loss_{Bayes-IPW}$ to   $loss_{opt}$ as follows:
\begin{align}
  \!\!\! & loss_{opt} = \!  \!\!\!\!\!\!\!   \sum_{u, x_i, x_j,  i,   j, \atop c_i >  c_j,c_j>0}  \!\!\!\!\!\!\!\!\!  \frac {L_{ij} m_{ij} | \Delta Z_{ij}| } {\theta_i \theta_j} \!  
  +  \!\!\!\!\!\!\!\!\!\!   \sum_{u, x_i, x_j,  i,   j, \atop  c_i > c_j,c_j=0}  \!\!\!\!\!\!\!  \frac {L_{ij} m_{ij}h_{ij} | \Delta Z_{ij}| } {\theta_i \theta_j} \!,  \!\!
\end{align}
where $\Delta Z_{ij}$ is the difference between evaluation metrics based on the biased synthesized label if item $x_i$ and $x_j$ are swapped in the ranking list.

Thus, if we can estimate the parameters of $\theta_i$, $\theta_i^-$, $\epsilon_{ij}^+$ and $\epsilon_{ij}^-$, we can learn an unbiased ranker $\hat{f}_{unbiased}$ through minimizing the loss function. In this paper, we employ a regression-based Expectation-Maximization (EM) method \cite{wang2018position} to estimate the parameters. The estimation process would be described in the next section.

\begin{figure*}[!htbp]
\setlength{\belowcaptionskip}{-0.5cm}
\setlength{\abovecaptionskip}{-0.05cm} 
	\centering
	\includegraphics[width=0.94\textwidth, trim=50 533 0 70,clip]{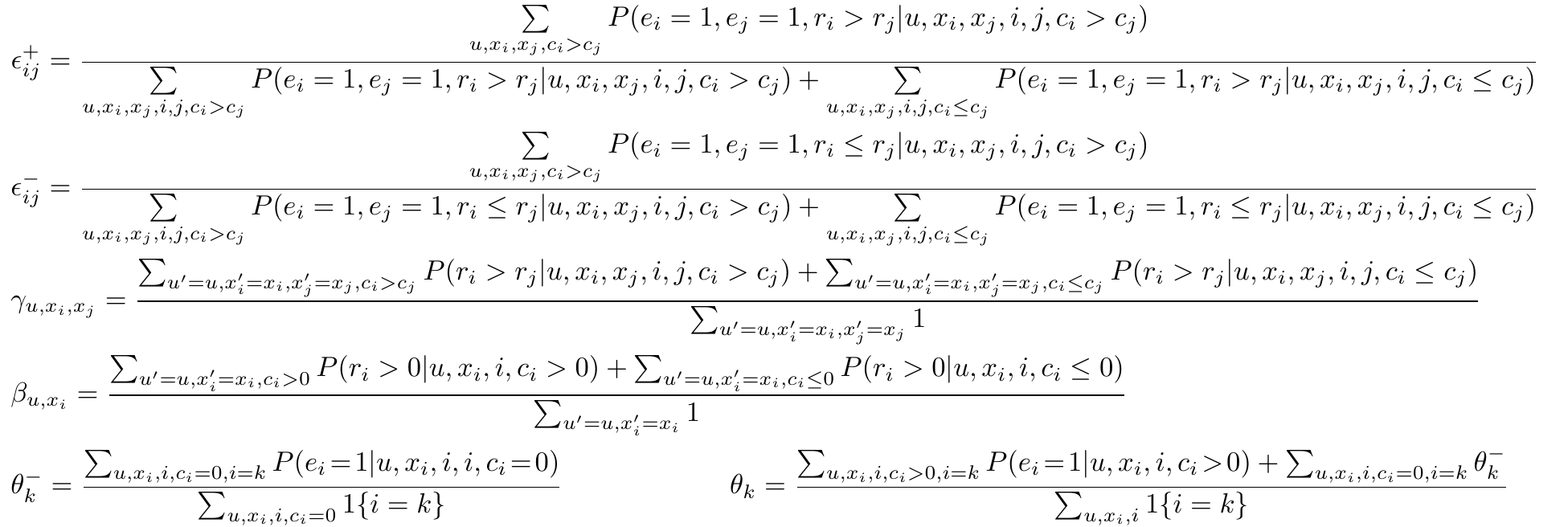}
	\caption{Formulas in Maximization Step}
	\label{mstep}
\end{figure*}
\vspace{-0.3cm}

\section{Estimation via Expectation-Maximization}

Expectation-Maximization (EM) \cite{moon1996expectation} is an iterative method to find maximum likelihood estimates of parameters. In this paper, we extend the previously proposed regression-based EM method \cite{wang2018position} to the pairwise setting to estimate parameters. In the procedure of regression-based EM algorithm, parameters are estimated by iterating over the Expectation steps and Maximization steps until convergence. In this section, we illustrate the process of the Expectation step and the Maximization step respectively.

\section{Expectation Step}
In the Expectation step, we need to estimate the distribution of hidden variable $e_i$, $e_j$ and relevance pair $(r_i,r_j)$ given parameters $\theta_i$, $\epsilon_{ij}^+$ and $\epsilon_{ij}^-$ and $\gamma_{u,x_i,x_j}$. To achieve the goal, we first calculate the joint probability of all hidden variables shown in \autoref{estep} , where all formulas \footnote{The detailed derivation of formulas in \autoref{estep} and \autoref{mstep} is presented in the supplementary material.} follow directly from Bayes rules. For instance, in the first equation, we have:
\begin{align}
& P(e_i=1,e_j=1,r_i>r_j|u,x_i,x_j,i,j,c_i>c_j)  \nonumber  \\
& = \frac {P(c_i \! > \! c_j|e_i \! = \! 1,e_j \! = \! 1,r_i \! > \!  r_j) P(e_i \! = \! 1,e_j \! = \! 1,r_i \! > \! r_j)} {P(c_i \! > \! c_j|u,x_i,x_j,i,j)}  \nonumber  \\
& = \frac { \theta_i \theta_j \epsilon_{ij}^+ \gamma_{u,x_i,x_j} } { \theta_i \theta_j  ( \epsilon_{ij}^+ \gamma_{u,x_i,x_j} \!\!\! + \! \epsilon_{ij}^-(1 \! - \! \gamma_{u,x_i,x_j})\!) \!\! + \! \theta_i (\! 1 \!\! - \! \theta_j\!) \beta_{u,x_i}}
\end{align}

It is worth noting that we estimate hidden variables $e_i$, $e_j$ and true relevance $r_i>0$ in the pointwise fashion, which is shown in last 4 equations in \autoref{estep}, as they only depend on a single item. After calculating the joint probability, we can then calculate the marginal probability of hidden variables, which will be used in the Maximization step. For instance, the marginal probability $P(e_i=1|c_i=0)$  equals to the sum of $P(e_i=1,r_i>0 | c_i=0)$ and $P(e_i=1,r_i \leq 0 | c_i=0)$.

\subsection{Maximization Step}
In the maximization step, all parameters are updated to minimize the loss function, given the training samples and the posterior probabilities from the Expectation step. All formulas for parameter updating are presented in \autoref{mstep}.
In the procedure of standard EM algorithm, $\gamma_{u,x_i,x_j}$ and $\beta_{u,x_i}$  are updated following the $3rd$ equation and $4th$  equation respectively, which requires that  $(u,x_i,x_j)$ should repeat and be shown in different position pairs. As samples of user-item-pair $(u,x_i,x_j)$ are highly sparse and synthesized label pairs $(c_i,c_j)$ are noisy in recommendation scenario, it is extremely difficult to minimize loss through free parameters $\gamma_{u,x_i,x_j}$ and $\beta_{u,x_i}$. To address the problem, we apply the regression-based EM \cite{wang2018position} algorithm to estimate parameters of $\gamma_{u,x_i,x_j}$ and $\beta_{u,x_i}$ via learning a regression function. 
Specifically, we assume that there are feature vector $X_{u,x_i,x_j}$  and $X_{u,x_i}$  representing the sample of user-item-pair $(u,x_i,x_j)$ and user-item $(u,x_i)$ respectively, we use function $g$ to compute the relevance preference $\gamma_{u,x_i,x_j} \!\!= \! g(X_{u,x_i,x_j })$ and function $h$ to compute absolute relevance $\beta_{u,x_i } \! = \! h(X_{u,x_i })$. Thus, we aim to find regression functions $g(X)$ and  $h(X)$ in maximization step to minimize loss function, given training samples and the distributions of hidden variables calculated in the Expectation step. For instance, we can regress the feature vector  $X_{u,x_i,x_j}$ to the probability $P( r_i \!\! > \!\! r_j|u,x_i,x_j,i,j,c_i,c_j )$. Similar to \cite{wang2018position}, we convert such a regression problem to a classification problem based on Bernoulli sampling. Specifically, we sample a binary relevance pair label $\gamma \in \{0,1\}$ and a binary relevance label $\beta \! \in \! \{0,1\}$ according to $P( r_i \!> \! r_j|u,x_i,x_j,i,j,c_i,c_j )$ and $P( r_i \!>0|u,x_i,i,c_i )$ respectively. Then we can adopt classification models to learn $g(X)$ and $h(X)$ based on training set $\{X,\gamma\}$ and $\{X,\beta\}$. It is flexible to use any classification model for $g(X)$ and $h(X)$ and we choose the commonly used DNN model in this paper.

	
Note that the original EM algorithm updates parameters in the Maximization step through calculation on the whole data set, which imposes challenges for industrial recommender systems with large-scale training data. To address the problem, we employ the mini-batch EM following the idea of online EM \cite{cappe2009line}. Accordingly, we calculate the formulas in Figure 2 based on data in one single mini-batch, and update parameters incrementally in each batch as follows:
\begin{equation}
\epsilon_{ij}^+ = \epsilon_{ij}^+ \times (1-\alpha) + \widehat {\epsilon_{ij}^+} \times \alpha
\end{equation}
where $\alpha$ is the scheduled learning rate of $\epsilon_{ij}^+$, $\widehat {\epsilon_{ij}^+} $ is the estimation of $\epsilon_{ij}^+$ in the current batch.

\begin{table*}
\setlength\tabcolsep{4pt}
\setlength{\belowcaptionskip}{-0.5cm}
\begin{tabular}{llcccccc}
\hline
& & \multicolumn{3}{c}{Yahoo Dataset } &  \multicolumn{3}{c}{MSLR-Web30K} \\
&& NDCG@3 & NDCG@5 & NDCG@10 & NDCG@3 & NDCG@5 & NDCG@10 \\
\hline

\multirow{2}*{\shortstack{Relevance \\ Label }} & Upper Bound (Pairwise) & 0.6942  & 0.7164 & 0.7630 & 0.4018 & 0.4092 & 0.4313 \\
 & Upper Bound (Pointwise) & 0.6936 & 0.7161 & 0.7628 & 0.3929 & 0.4011 & 0.4244 \\
  \hline
 \multirow{6}*{\shortstack{Categorical \\ Label}}  
& Our Method ($loss_{opt}$) &  0.6895 & \textbf{0.7115} & \textbf{0.7589} & \textbf{0.3811} & \textbf{0.3899} & \textbf{0.4132} \\
~ & Our Method ($loss_{Bayes-IPW}$) &  \textbf{0.6900} & \textbf{0.7116} & \textbf{0.7589} & 0.3800 & 0.3888 & 0.4125 \\
~ & Our Method ($loss_{IPW}$) & 0.6860 & 0.7079 & 0.7557 & 0.3784 & 0.3882 & 0.4119\\
~ & DLA \cite{ai2018unbiased} & 0.6774 & 0.7001 & 0.7494 & 0.3792 & 0.3870 & 0.4102 \\
~ & Regression EM \cite{wang2018position} & 0.6893 & 0.7108 & 0.7577 & 0.3707 & 0.3766 & 0.3977 \\
~ & Pairwise Debiasing \cite{hu2019unbiased} & 0.6600 & 0.6847 & 0.7365 & 0.3481 & 0.3623 & 0.3912 \\ 
~ & Lower Bound (Pairwise) & 0.6541 & 0.6790 & 0.7320 & 0.3433 & 0.3579 & 0.3874 \\
~ & Lower Bound (Pointwise) & 0.6530 & 0.6770 & 0.7285 & 0.3407 & 0.3522 & 0.3790 \\
\hline
\multirow{6}*{\shortstack{Continuous \\ Label}} & Our Method ($loss_{opt}$) &  \textbf{0.6877} & \textbf{0.7095} & \textbf{0.7568} & \textbf{0.3886} & \textbf{0.3963} & \textbf{0.4189} \\
~ & Baseline V1 &   {0.6630} &  {0.6858} &  {0.7351} &  {0.3387} &  {0.3471} &  {0.3711} \\
~ & Baseline V2  & 0.6712 & 0.6936 & 0.7429 & 0.3556 & 0.3653 & 0.3892 \\
~ & Baseline V3 & 0.6740 & 0.6963 & 0.7452 & 0.3596 & 0.3685 & 0.3916 \\ 
~ & Baseline V4 & 0.6713 & 0.6936 & 0.7423 & 0.3412 & 0.3494 & 0.3729 \\
~ & Lower Bound (Pairwise) & 0.6547 & 0.6800 & 0.7329 & 0.3458 & 0.3601 & 0.3888 \\
~ & Lower Bound (Pointwise) & 0.6663 & 0.6890 & 0.7384 & 0.3557 & 0.3652 & 0.3901 \\
\hline
\end{tabular}
\caption{Experiment Results on Yahoo Dataset and MSLR-Web30K}
\label{table1}
\end{table*}

\section{Experiments}
In this section, we conduct offline experiments on benchmark datasets and online A/B testing in one of the world's largest content recommendation platforms to evaluate the performance of the proposed method.

\subsection{Experiments on Benchmark Dataset}
To evaluate the effectiveness of the proposed method, we conduct experiments on two of the largest public LTR benchmark datasets, i.e., Yahoo! LETOR set 1\footnote{https://webscope.sandbox.yahoo.com/catalog.php?datatype=c} and MSLR-Web30K \footnote{https://www.microsoft.com/en-us/research/project/mslr/}, which both consists of multiple query document pairs represented by feature vectors and 5-level relevance labels. As there is no user action in the datasets, we consider each query as a user request and generate the click and dwell-time labels to simulate key tasks in recommender systems.  We follow the same data split of training, validation, and testing in the datasets and conduct experiments on both binary click labels and continuous synthesized labels to evaluate the performance of our method comprehensively.

\subsubsection{Data Simulation.}
In the experiment, we first sample clicks following the same procedure as \cite{ai2018unbiased}. Then we sample continuous dwell-time labels based on sampled click data as only clicked samples have non-zero dwell-time in practice, which is formulated as:
\begin{equation}
d_{u,x_i,i} = c_{u,x_i} \delta_i \omega_{u,x_i}, \nonumber
\end{equation}
where $c_{u,x_i}$ is the sampled click label, $\delta_i$ is the position bias of dwell-time at position $i$ and $\omega_{u,x_i}$ is the unbiased dwell-time related to the relevance. As dwell-time is continuous, we assume $\delta_i$ and $\omega_{u,x_i}$ follows the following normal distributions:

$\delta_i \! \sim \! N(\frac {2} {\sqrt{i+2}}, \frac {0.4} {\sqrt{i+2}}), \omega_{u,x_i} \! \sim \! N(\epsilon + (1-\epsilon)y, \frac {\sqrt{y} + \epsilon} {y_{max}+2}),  $
where $y \in [0,4]$ is the relevance label, $y_{max}$ is the max relevance of 4,  $\epsilon$ is the noise due to that irrelevant items may have non-zero dwell-time, which is set to 0.1 here.

\subsubsection{Experiments on Categorical Labels.}
We first perform experiments on the sampled binary click labels to compare our method with the SOTA unbiased LTR methods including DLA \cite{ai2018unbiased}, Regression EM \cite{wang2018position} and Pairwise Debiasing \cite{hu2019unbiased}. For comprehensive comparison, we train models with the unbiased relevance labels and the biased click labels without debiasing as upper bound and lower bound respectively, and adopt both pointwise learning and pairwise learning for lower bound and upper bound. Moreover, we remove the pairwise trust bias correction and loss refinement for direct metrics optimization from our method to explore their effectiveness. For each method, we adopt a  DNN model with [512, 256, 128] hidden units and ELU activation and tune hyper-parameters carefully. We train and evaluate each method 10 times and report the mean NDCG in \autoref{table1}. It is shown that our method outperforms all SOTA unbiased LTR methods in categorical labels. Furthermore, the pairwise trust bias correction achieves significant improvement on Yahoo Dataset and slightly better results on Web30K. Besides, direct optimization for ranking metrics exhibits better performance on Web30K but similar performance on Yahoo dataset as we can only compute $\Delta Z_{ij}$ based on the noisy click signal. On the whole, our method achieves the best performance for unbiased LTR.

\subsubsection{Experiments on Continuous Labels.}
Then we generate the synthesized label following Equation 6 with summation combination function and conduct experiments to evaluate the performance of our method on the setting similar to the typical recommendation scenarios. The lower bound here is trained with the synthesized labels without debiasing. Note that the SOTA industrial recommender systems train multiple pointwise models to predict multiple objectives respectively and employ PAL \cite{guo2019pal} or sum-based shallow tower \cite{zhao2019recommending} for debiasing. Moreover, PAL can only work for categorical labels due to its decomposition. So we consider the following baseline methods:
\begin{itemize}
    \item Baseline V1: We train a ranker with the synthesized label and adopt the sum-based shallow tower for debiasing.
    \item Baseline V2: We train the click and dwell-time task respectively and adopt sum-based shallow tower for debiasing on the click task.
    \item Baseline V3: We train the click and dwell-time task respectively and adopt PAL for debiasing on the click task.
    \item Baseline V4: We train the click and dwell-time task respectively and adopt PAL on the click task and sum-based shallow tower on the dwell-time task for debiasing.
\end{itemize}


For each task, we adopt a DNN model with [512, 256, 128] hidden units, ELU activation, and cross-entropy loss for classification task and MSE loss for regression task. For Baseline V2, V3 and V4, we sum the predicted click ratio and dwell-time for ranking. According to the mean NDCG over 10 runs shown in \autoref{table1}, our method outperforms all baseline methods significantly in both datasets. One can see that pairwise learning performs worse than pointwise learning with the biased synthesized label while performs much better on relevance label, which indicates that debiasing is more important for pairwise learning in recommendation scenarios. Moreover, employing shallow tower on dwell-time harms the performance significantly, which shows that shallow tower cannot work well for continuous labels. In contrast, our probabilistic graphical based method has a clear separation of bias and relevance and achieves superior performance on debiasing and preference learning for different types of labels and applications.

\subsection{Online A/B Testing}
We have deployed our method in one of the world's content recommendation platforms and conducted online A/B testing for 7 days. The live metric and synthesized label are defined as the combination of multiple user actions (e.g. click, dwell-time, liking, and sharing). Comparing with the online baseline method similar to  Baseline V3 which models multiple user actions respectively and employs PAL on the click task for debiasing, the proposed method improves the live metric by 2.08\%, which demonstrates the effectiveness of our method in debiasing and preference learning in real-world applications.
\section{Conclusion}
In this paper, we propose a novel unbiased pairwise LTR method to model position-based examination bias and trust bias in the pairwise fashion for better bias correction and preference learning, which is the first unbiased LTR method working for both categorical and continuous labels. Offline experiment results on public benchmark datasets and online  A/B testing show significant and consistent improvements of the proposed method over SOTA ranking models in real-world recommender systems and SOTA unbiased LTR methods commonly used in search engines. Correcting for other types of bias besides position bias will be the focus of future work.

\clearpage
\bibliography{ref}

\end{document}